\documentclass[11pt,a4paper]{article}
\usepackage{jheppub} 

\usepackage{graphicx}
\usepackage{color}
\usepackage{amsmath}
\usepackage{amssymb}
\usepackage{amsthm}
\usepackage{epstopdf}
\usepackage{url}
\usepackage{hyperref}
\usepackage[footnotesize]{caption}
\begin{document}

  \title{Stringy Stability of Charged Dilaton Black Holes \\with Flat Event Horizon}
   \author[b,c]{Yen Chin Ong}
    
   \author[a,b,c,d]{Pisin Chen}
    
     \affiliation[a]{Graduate Institute of Astrophysics, National Taiwan University, Taipei, Taiwan 10617}
\affiliation[b]{Department of Physics, National Taiwan University, Taipei, Taiwan 10617}
\affiliation[c]{Leung Center for Cosmology and Particle Astrophysics, National Taiwan University, Taipei, Taiwan 10617}
\affiliation[d]{Kavli Institute for Particle Astrophysics and Cosmology, SLAC National Accelerator Laboratory, Stanford University, Stanford, CA 94305, U.S.A.}

 \emailAdd{d99244003@ntu.edu.tw}
  \emailAdd{pisinchen@phys.ntu.edu.tw}

\abstract{
Electrically charged black holes with flat event horizon in anti--de Sitter space have received much attention due to various applications in Anti--de Sitter/Conformal Field Theory (AdS/CFT) correspondence, from modeling the behavior of quark-gluon plasma to superconductor. Crucial to the physics on the dual field theory is the fact that when embedded in string theory, black holes in the bulk may become vulnerable to instability caused by brane pair-production. Since dilaton arises naturally in the context of string theory, we study the effect of coupling dilaton to Maxwell field on the stability of flat charged AdS black holes. In particular, we study the stability of Gao-Zhang black holes, which are locally asymptotically anti--de Sitter. We find that for dilaton coupling parameter $\alpha > 1$, flat black holes are stable against brane pair production, however for $0 \leq \alpha < 1$, the black holes eventually become unstable as the amount of electrical charges is increased. Such instability however, behaves somewhat differently from that of flat Reissner--Nordstr\"om black holes. In addition, we prove that the Seiberg-Witten action of charged dilaton AdS black hole of Gao-Zhang type with flat event horizon (at least in 5-dimension) is always logarithmically divergent at infinity for finite values of $\alpha$, and is finite and positive in the case $\alpha \to \infty$. We also comment on the robustness of our result for other charged dilaton black holes that are not of Gao-Zhang type.
}

\arxivnumber{1205.4398}

\maketitle
\flushbottom

\section{Introduction}

It is well known that since anti--de Sitter space does not satisfy the dominant energy condition, it admits black hole solutions with non-trivial horizon topology \cite{Birmingham}. Black hole solutions in anti--de Sitter space are of great importance due to various applications in AdS/CFT \cite{Maldacena, Witten1}. For example, Reissner--Nordstr\"om black holes provide a mean to understand quark--gluon plasma (QGP) \cite{Chamblin, McInnes1, McInnes2}. On the other hand, condensed matter systems such as superconductors \cite{Hartnoll, Horowitz, Herzog}, spin-model \cite{Gursoy} and strange metals \cite{Polchinski, Subir1, Subir2, Puletti} have also been explored via Anti--de Sitter/Condensed Matter Theory (AdS/CMT) correspondence [See \cite{Hartnoll2} and \cite{Sachdev}, as well as references therein, for an introductory review].  In addition, quantum hall effects have been studied via Anti--de Sitter/Quantum Hall Effect (AdS/QHE) correspondence \cite{Bayntun}. Since the applications to QGP, AdS/CMT and AdS/QHE make use of the well-tested conjecture that conformal field theory on $(d-1)$-dimensional boundary corresponds dually to black hole physics in $d$-dimensional bulk in which the strings are weakly coupled, calculations in the bulk can be done semiclassically. This provides a means to probe systems with strongly interacting electrons such as high $T_c$ cuprate superconductors, in which Landau's independent electron approximation breaks down. Unlike the case with QGP however, the study of condensed matter system typically involves black holes with scalar hair. 

For example, in a typical superconductor, a pair of electrons with opposite spin can bind to form a charged boson, which is known as a Cooper pair. Second order phase transition occurs below a critical temperature $T_c$, and these bosons condense, leading to the divergence of direct current conductivity. The gravity dual to it should then consist of a black hole with electrical charges (which is related to the temperature of the black hole, and hence dual to the temperature of the superconductor) that eventually becomes unstable and develops scalar hair (which corresponds to the boson condensate). A possible choice for such a gravity dual is thus a Reissner--Nordstr\"om black hole that becomes unstable at temperature $T < T_c$ and subsequently turns ``hairy'' . Since dilaton occurs naturally in string theory, it is a natural choice (though not the only one) for the relevant scalar in this context. Embedding the black hole into string theory however, exposes it to the menace of Seiberg-Witten instability \cite{SeibergWitten, Kleban}, which is induced by uncontrolled brane pair production, analogous to the Schwinger mechanism of vacuum polarization \cite{Schwinger}. Indeed, it has been shown that Reissner--Nordstr\"om black holes do ``universally'' become unstable in the Seiberg-Witten sense for all spacetime dimensions above 3 \cite{McInnes3}. It is therefore interesting to study whether charged dilaton black holes are also vulnerable to Seiberg-Witten instability. 

Since applications in AdS/CMT are usually concerned with field theory on locally flat spacetime, in this work we will focus on studying black holes with flat event horizon. We first review various properties of charged dilaton black holes in Section 2, followed by the analysis of the Seiberg-Witten stability of charged dilaton AdS black hole of Gao-Zhang type in Section 3. Finally we conclude with some discussions, including some remarks on the dual field theory in Section 4, and end with some remarks on the robustness of our results.

\section{Charged Dilaton Black Holes}

\subsection{The Garfinkle--Horowitz--Strominger Black Hole}

We first review some properties of charged dilaton black holes, starting with those with spherical horizon in $(3+1)$-dimensional asymptotically flat spacetime as these are more readily compared to the more familiar Reissner--Nordstr\"om solution. Furthermore some crucial properties of these black holes are retained even in the case of flat black holes in anti-de Sitter space, which is the focus of our work.  

We recall that the four-dimensional low-energy Lagrangian obtained from string theory is 
\begin{equation}
S= \int d^4x \sqrt{-g} \left[-R + 2(\nabla \phi)^2 + e^{-2\phi}F^2\right]
\end{equation}
where $F=F_{\mu\nu}F^{\mu\nu}$ with $F_{\mu\nu}$ being the Maxwell field associated with a $U(1)$ subgroup of $E_8 \times E_8$ or $\text{Spin}(32)/\mathbb{Z}_2$, and $\phi$ is the dilaton, a scalar field that couples to the Maxwell field. The dilaton field has some value $\phi_0$ at infinity, which we have set to be zero (or by absorbing a factor of $e^{-\phi_0}$ into the normalization of the Maxwell field).

Garfinkle, Horowitz and Strominger obtained in their landmark paper \cite{GHS} (also see \cite{Gibbons} and \cite{Maeda}) the charged black hole solution to this action. Its metric is
\begin{equation}
g(\text{GHS}) = -\left(1-\frac{2M}{r}\right)dt^2 + \left(1-\frac{2M}{r}\right)^{-1}dr^2 + r\left(r-\frac{Q^2}{M}\right)d\Omega^2
\end{equation}
where $d\Omega^2$ is the standard metric on 2-sphere. We assume $Q>0$ throughout this work for simplicity. 

Recall that Reissner--Nordstr\"om black hole has two horizons, which merge into one, and its Hawking temperature smoothly goes to zero, as the extremal limit is reached. This prevents naked singularity from ever forming. However, since the extremal black hole has nonzero area $A$, it has nonzero entropy in accordance with the usual $S=A/4$ area law for Bekenstein-Hawking entropy \cite{Bekenstein1, Bekenstein2, Hawking} (However, there are arguments in favor of extremal Reissner--Nordstr\"om black hole having zero entropy despite appearances, see. e.g. \cite{HHR, Ghosh, Shahar, BM, SJR, EC}). We remark that the scaling law $S=A/4$ is strictly true only in the weak gravity regime, or for large black holes. In the small mass limit $M \to 0$, the relationship becomes complicated due to UV modification of uncertainty principle, also known as \emph{Generalized Uncertainty Principle} (GUP) \cite{AJC}. Therefore, in general vanishing area is \emph{not} a precondition for vanishing entropy.

The GHS black hole behaves very differently since its event horizon remains at $r=2M$ as more electrical charges are dropped into the black hole; yet at the same time its \emph{area} decreases. In fact at extremal charge $Q=\sqrt{2}M$, the black hole has \emph{vanishing} area and hence \emph{zero} entropy. The Hawking temperature of GHS black hole is independent of the amount of electrical charges \cite{GHS}: $T = (8\pi M)^{-1}$. The extremal GHS black hole is not a black hole in the ordinary sense since its area has become degenerate and \emph{singular}: it is in fact a naked singularity. Unlike singularity of Reissner--Nordstr\"om which is timelike, this singularity is in fact \emph{null}, i.e. outward-directed radial null geodesics cannot hit it. For a detailed study of null geodesics of GHS black hole, see \cite{Fernando}. 

We remark that strings do not couple directly to the physical metric, but rather to the conformally related string metric $g(\text{string})=e^{2\phi}g(\text{GHS})$ (See, e.g. \cite{GHorowitz}). If the black hole is magnetically charged, then the string metric is:
\begin{equation}
g(\text{string})=-\left(\frac{1-\frac{2M}{r}}{1-\frac{Q^2}{Mr}}\right)dt^2 + \frac{dr^2}{\left(1-\frac{2M}{r}\right)\left(1-\frac{Q^2}{Mr}\right)} + r^2d\Omega^2.
\end{equation} 
Note that at the extremal radius $r=\sqrt{2}M$, the string coupling diverges:
\begin{equation}\label{stringframe}
e^{2\phi} = \frac{1}{1-\frac{Q^2}{Mr}} \rightarrow \infty.
\end{equation}
This leads to blowing up of string frame curvature at the horizon, which suggests that higher derivative corrections should be considered near to the extremal horizon. Therefore any naive semi-classical calculation and interpretation on the properties of extremal dilaton black holes should be taken with healthy dose of skepticism. 
However if the black hole is electrically charged, then the string metric is different:
\begin{equation}
g(\text{string})=-\left({1-\frac{2M}{r}}\right)\left(1-\frac{Q^2}{Mr}\right)dt^2 + \frac{\left(1-\frac{Q^2}{Mr}\right)}{\left(1-\frac{2M}{r}\right)}dr^2 + r^2\left(1-\frac{Q^2}{Mr}\right)^2d\Omega^2.
\end{equation}
Under the electromagnetic duality, $\phi \to -\phi$ as we map electrically charge black hole solution to a magnetically charged one, thus the string coupling for the electrical charged hole remains small near the an extremal horizon.  In this paper we are only concerned with the electrically charged holes.

Lagrangian with generalized dilaton coupling of the form
\begin{equation}
S= \int d^4x \sqrt{-g} \left[-R + 2(\nabla \phi)^2 + e^{-2\alpha\phi}F^2\right], ~\alpha \geq 0,
\end{equation}
has been considered in \cite{GHS} and \cite{Maeda}. This type of action can arise in string theory if $F$ is a Maxwell field arising in the compactification process, or in the
type IIA string theory \cite{GHS}. The case of $\alpha = \sqrt{3}$ corresponds to the 4-dimensional effective model reduced from the 5-dimensional Kaluza-Klein theory. Furthermore, the case $\alpha = \sqrt{p/(p+2)}, ~p=0,1,2,...$ can arise from toral $T^p$ compactification of static truncation of $(4+p)-$dimensional Einstein-Maxwell theory \cite{Townsend}.

The black hole solution, known as the Garfinkle--Horne black hole \cite{Horne}, takes the form
\begin{equation}
g(\text{GH}) = -\left(1-\frac{r_+}{r}\right)\left(1-\frac{r_-}{r}\right)^{\frac{1-\alpha^2}{1+\alpha^2}} dt^2 + \left(1-\frac{r_+}{r}\right)^{-1}\left(1-\frac{r_-}{r}\right)^{\frac{\alpha^2-1}{\alpha^2 +1}}dr^2 + r^2\left(1-\frac{r_-}{r}\right)^{\frac{2\alpha}{1+\alpha^2}}d\Omega^2,
\end{equation}
where $r_+$ and $r_-$  denote the event horizon and the inner horizon respectively. The GH solution reduces to Reissner--Nordstr\"om solution for $\alpha=0$ and to GHS solution for $\alpha=1$ (In general, constant dilaton reduces to the Brans-Dicke-Maxwell theory with Brans-Dicke constant $\omega=-1$. In the case $\alpha=0$ this further reduces to Einstein-Maxwell theory due to supersymmetry \cite{Nozawa}). Like the GHS black hole, the more general GH black hole has vanishing area and thus vanishing entropy in the extremal limit (See, however, \cite{Cox}, for the argument that there exist values of $\alpha$ in which the extremal limit does not become naked singularity, but instead evolve from membranes to strings). Its Hawking temperature is
\begin{equation}
T=\frac{1}{4\pi r_+} \left(\frac{r_+-r_-}{r_+}\right)^{\frac{1-\alpha^2}{1+\alpha^2}},
\end{equation}  
which goes to zero as the black hole becomes extremal for $\alpha < 1$, remains finite and constant for $\alpha=1$. It is formally infinite for $\alpha >1$, yet it turns out that the black hole develops a \emph{mass gap} which is of the same order as the temperature, so one might hope that Hawking radiation will be effectively shut off in the extremal limit\cite{HW}. However, it was later shown that despite the competing process from the effective potential, the emission rate for $\alpha>1$ black holes \emph{does} diverge in the extremal limit, and so near-extremal $\alpha > 1$ black holes are likely to be unstable \cite{Koga}.

\subsection{Charged Dilaton Black Holes in Anti-de Sitter Space}

Since AdS/CFT has become a powerful tool in the study of strongly-coupled field theories, it is only natural for one to seek AdS generalizations of charged dilaton black holes. The effort was not straightforward. It was first found that with the exception of pure cosmological constant, no dilaton-de Sitter or dilaton-anti-de Sitter black hole solution exists if there is only one Liouville-type dilaton potential \cite{Poletti1,Poletti2,Mignemi}. Subsequently, black hole solutions which are neither asymptotically flat nor asymptotically (A)dS were discovered \cite{Chan, Clement, Ahmad}, and thus do not apply directly to AdS/CFT. Finally, Gao and Zhang \cite{GZ} successfully obtained static charged dilaton black hole solutions which are asymptotically AdS, by using a combination of \emph{three} Liouville-type dilaton potentials. 

The Einstein-Maxwell-Dilaton action in $n$-dimensional spacetime is 
\begin{equation}
S=-\frac{1}{16\pi}\int d^nx \sqrt{-g} \left[R -\frac{4}{n-2}(\nabla \phi)^2 - V(\phi) - e^{-\frac{4\alpha \phi}{n-2}}F^2\right], ~\alpha \geq 0,
\end{equation} 
where the dilaton potential is expressed in terms of the dilaton field and its coupling to the cosmological constant:

\begin{flalign}\label{V}
V(\phi) &= \frac{\Lambda}{3(n-3+\alpha^2)^2}\Bigg[-\alpha^2(n-2)(n^2-n\alpha^2-6n+\alpha^2+9)\text{exp}\Bigg(\frac{-4(n-3)\phi}{\alpha(n-2)}\Bigg) \notag \\ &~+ (n-2)(n-3)^2(n-1-\alpha^2)\text{exp}\Bigg(\frac{4\alpha \phi}{n-2}\Bigg) \notag \\ &~+ 4\alpha^2 (n-3)(n-2)^2 \text{exp}\Bigg(\frac{-2\phi (n-3-\alpha^2)}{(n-2)\alpha}\Bigg)\Bigg].
\end{flalign}
We have again set the asymptotic value of the dilaton field to zero. The topological black hole solutions take the form \cite{GZ}
\begin{equation}\label{Metric}
ds^2 = - g_{tt}dt^2 + g_{rr}dr^2 + \left[f(r)\right]^2 d\Omega^2_{k,n-2},
\end{equation}
where 
\begin{equation}
g_{tt} = \left[k-\left(\frac{c}{r}\right)^{n-3}\right]\left[1-\left(\frac{b}{r}\right)^{n-3}\right]^{1-\gamma(n-3)} - \frac{2\Lambda}{(n-1)(n-2)}r^2 \left[1-\left(\frac{b}{r}\right)^{n-3}\right]^\gamma,
\end{equation}
\begin{equation}
g_{rr} = \left[g_{tt}\right]^{-1}\left[1-\left(\frac{b}{r}\right)^{n-3}\right]^{-\gamma(n-4)},
\end{equation}
and
\begin{equation}
\left[f(r)\right]^2 = r^2 \left[1-\left(\frac{b}{r}\right)^{n-3}\right]^\gamma,
\end{equation}
with
\begin{equation}
\gamma = \frac{2\alpha^2}{(n-3)(n-3+\alpha^2)}.
\end{equation}
The cosmological constant is related to spacetime dimension $n$ by
\begin{equation}
\Lambda = -\frac{(n-1)(n-2)}{2L^2},
\end{equation}
where $L$ denotes the AdS length scale. We will choose a unit in which $L=1$ for simplicity. Note that Gao and Zhang use the effective cosmological constant $\Lambda_{\text{eff}} = -3/L^2$ that is independent of dimensionality. This is, however, equivalent to our notation, which follows that of \cite{HSD}. The event horizon is a compact $(n-2)$-dimensional Riemannian manifold of constant curvature $k=-1,0,+1$, and $d\Omega^2_{k,n-2}$ is the metric on this space. We shall denote the area of this space (for $r=1$) by $\Gamma_{n-2}$.

We note that in the special case $n=4$, $\Lambda=0$, and $k=1$ the integration parameters $c$ and $b$ reduce to the event horizon $r_+$ and inner horizon $r_-$ respectively of the GH black hole solution, but they are in general \emph{not} the horizons of the black hole. Indeed, explicit calculation for $(4+1)$-dimensional flat black hole has been performed in the $\alpha=1$ case \cite{yenchin}, in which the event horizon was found to satisfy $r_+ = b^{1/2}$. As shown in \cite{HSD}, the mass of the black hole can be obtained following the method of Brown and York \cite{BY} as  
\begin{equation}\label{Mass}
M = \frac{\Gamma_{n-2}}{16\pi} \left[c^{n-3} + k\left(\frac{n-3-\alpha^2}{n-3+\alpha^2}\right)b^{n-3} \right].
\end{equation}
The electrical charge parameter is
\begin{equation}
q^2 = \frac{(n-2)(n-3)^2}{2(n-3+\alpha^2)}(bc)^{n-3},
\end{equation}
and the ADM charge is 
\begin{equation}\label{Charge}
Q = \frac{1}{4\pi} \lim_{r\to \infty} \int d^{n-2}x \sqrt{-g} F_{tr} = \frac{\Gamma_{n-2}}{4\pi}q,
\end{equation}
where 
\begin{equation}
F_{tr} = \frac{q}{r^{n-2}}e^{\frac{4\alpha \phi}{n-2}}\left[1-\left(\frac{b}{r}\right)^{n-3}\right]^{\frac{\gamma}{n-3}}
\end{equation}
is the only non-vanishing component of the Maxwell field strength tensor $F$. Furthermore, the dilaton satisfies
\begin{equation}
e^{2\phi} = \left[1-\left(\frac{b}{r}\right)^{n-3}\right]^{-(n-2)\sqrt{\gamma(2+(3-n)\gamma)}/2}.
\end{equation} 

For our subsequent analysis, by charged dilaton black holes we always mean Gao- Zhang black holes unless otherwise stated. Since the Gao-Zhang solution is obtained from a very special kind of potential given by Eq.(2.9), one may wonder what happens if the charged dilaton black holes are not of Gao-Zhang type? We will address this question in Section 4.

For concreteness, we will focus on $(4+1)$-dimensional flat black holes from here onwards, that is $L=1,~n=5,~k=0$. The event horizon is a 3-dimensional flat compact Riemannian manifold. These are Conway's ``Platycosms'' \cite{Conway}. There are in fact a variety of such manifolds, six to be exact if we only consider orientable case. Dilaton black holes with flat event horizon were also previously studied in e.g. \cite{rg1} and \cite{rg2}.

The simplest case is a cubical 3-torus, which can be obtained by topologically identifying opposite sides of a cube. For this cubic torus, seen as the product $S^1 \times S^1 \times S^1$, where each copy of $S^1$ has length $2\pi K$, its area is simply $\Gamma=8\pi^3 K^3$. We then use this relation to \emph{define} area of \emph{any} 3-dimensional flat compact Riemannian manifold. In other words, $K$ now serves as the \emph{area parameter} that measures the relative size of the space, as it deviates away from being a cubic torus. This follows the notations of \cite{McInnes4}. The mass of the charged dilaton black hole is then, by Eq.(\ref{Mass}),  
\begin{equation}
M=\frac{3}{2}\pi^2 K^3 c^2,
\end{equation}
and the charge is, by Eq.(\ref{Charge}),
\begin{equation}
Q^2 = \frac{2}{2+\alpha^2}\left(48\pi^5 K^6b^2 c^2\right).
\end{equation}
Consequently, 
\begin{equation}
b^2 = \frac{Q^2(2+\alpha^2)}{96\pi^5 K^6 c^2} = \frac{Q^2(2+\alpha^2)}{96 \pi^5 K^6 \left(\frac{2M}{3\pi^2K^3}\right)} = \frac{Q^2(2+\alpha^2)}{64\pi^3 K^3 M}.
\end{equation}
The coefficients of our metric [Eq.(\ref{Metric})] reduce to 
\begin{equation}\label{gtt}
g_{tt} = -\left(\frac{c}{r}\right)^2\left[1-\left(\frac{b}{r}\right)^2\right]^{\frac{2-\alpha^2}{2+\alpha^2}} + r^2\left[1-\left(\frac{b}{r}\right)^2\right]^{\frac{\alpha^2}{2+\alpha^2}},
\end{equation}
\begin{equation}
g_{rr} = \left[g_{tt}\right]^{-1} \left[1-\left(\frac{b}{r}\right)^2\right]^{-\frac{\alpha^2}{2+\alpha^2}},
\end{equation}
and 
\begin{equation}
\left[f(r)\right]^2 = r^2\left[1-\left(\frac{b}{r}\right)^2\right]^{\frac{\alpha^2}{2+\alpha^2}}.
\end{equation}

In general, the solution is very complicated, and it is not possible to explicitly solve for the event horizon in terms of the metric parameters for general $\alpha$. This task is significantly easier in the case $\alpha=1$, since $g_{tt}$ factorizes into
\begin{equation}
g_{tt}(\alpha=1) = \left[r^2 - \left(\frac{c}{r}\right)^2\right]\left[1-\left(\frac{b}{r}\right)^2\right]^{\frac{1}{3}}.
\end{equation} 
The event horizon and the inner horizon in that case are, respectively \cite{yenchin}, 
\begin{equation}
r_+ = c^{1/2} = \left(\frac{2M}{3\pi^2 K^3}\right)^{\frac{1}{4}}
\end{equation}
and
\begin{equation}
r_- = b = \left(\frac{3Q^2}{64M\pi^3 K^3}\right)^{\frac{1}{4}}.
\end{equation}
It is noted that the position of the event horizon for charged dilaton AdS black hole in the case $\alpha=1$ is independent of the electrical charge, just like its asymptotically flat GHS cousin. Furthermore, the horizon becomes singular at extremal charge $Q_E \approx 13.11 (KM)^{3/4}$, which exceeds that of the extremal charge of flat Reissner--Nordstr\"om counterpart, $Q_E (\text{RN}) \approx 9.96 (KM)^{3/4}$ \cite{McInnes1}. This is similar to the behavior of $(3+1)$-dimensional GHS black hole having extremal charge $\sqrt{2}M$, which is larger than that of (spherical) Reissner--Nordstr\"om black hole. Also, for general $\alpha \neq 0$, the Kretschmann scalar $R_{\mu\nu\lambda\rho}R^{\mu\nu\lambda\rho}$ as well as the Ricci scalar diverge at $r=b$ \cite{HSD}, so that $r=b$ is a (null) curvature singularity \cite{HSD}.

\section{Seiberg--Witten Stability of Charged Dilaton Black Holes}

There is, by now, an extensive literature on the applications of charged dilaton black holes in the context of AdS/CFT, see, e.g. \cite{Goldstein, CMChen, CGKKM, Cadoni, GKM, LKNT, GK, JPW, LMZ}. Gubser and Rocha \cite{GR} for example, argued that charged dilaton black holes or a relative with similar behavior could be dual to Fermi liquid. 

In the study of QGP, since the plasma cannot be \emph{arbitrarily} cold, it only makes sense that its gravity dual, the flat Reissner--Nordstr\"om black holes, cannot be allowed to approach extremal limit (which has vanishing temperature) arbitrarily close \cite{McInnes1, McInnes2}. This is achieved via Seiberg--Witten instability, which refers to uncontrolled nucleation of branes in spacetimes with Euclidean version of a particular kind. Specifically, in the context of Type IIB string theory on $\text{AdS}_5 \times S^5$, the \emph{Seiberg--Witten action} is defined on the Euclidean spacetime obtained after Wick rotation by 
\begin{equation}\label{SWaction}
S_{\text{SW}}=\Theta (\text{Brane Area}) - \mu (\text{Volume Enclosed by Brane}),
\end{equation}
where $\Theta$ is related to the tension of the brane and $\mu$ relates to the charge enclosed by the brane due to the background antisymmetric tensor field. This brane is essentially a probe that allows us to study the background fields and geometry of the bulk. The probe brane is assumed not to disturb the bulk geometry and background fields. Seiberg and Witten have shown very generically that non-perturbative instability occurs when the action becomes negative due to uncontrolled 3-brane productions. Brane-anti-brane pairs are always spontaneously created from the AdS vacuum. In analogy to Schwinger effect in QED \cite{Schwinger}, the rate of brane-anti-brane pair production is proportional to $\exp(-S)$ where $S$ is the Seiberg--Witten action. Thus, if $S$ is negative, the AdS vacuum nucleates brane-anti-brane pairs at exponentially large rate instead of exponentially suppressed. This disturbs the background geometry so much that the spacetime is no longer described by the metric that we started with, i.e. the original spacetime is not stable if such brane-anti-brane production is exponentially enhanced due to the reservoir of negative action. Seiberg--Witten instability can occur, e.g. if the Seiberg--Witten action becomes negative at large $r$ near the boundary, which will happen if the boundary (conformal) metric has negative scalar curvature \cite{SeibergWitten}. To understand this picture in terms of brane and anti-brane dynamics in Lorentzian picture in more details, see \cite{Barbon}. 

Since zero-temperature limit can be indefinitely approached in the case of Fermi liquid as well as many superconductors, we would like to see that flat charged dilaton black holes are \emph{stable} against brane nucleation, at least for some strength of dilaton coupling $\alpha$. Of course, even this statement is in itself over-simplified, and we relegate further discussions to the final section. 

We remark that in many of these applications however, one usually only considers effective action of the form
\begin{equation}\label{Effective}
S=-\frac{1}{16\pi}\int d^nx \sqrt{-g} \left[R -\frac{4}{n-2}(\nabla \phi)^2 - V_0 e^{-\frac{4\delta \phi}{n-2}} - e^{-\frac{4\alpha \phi}{n-2}}F^2\right], ~\alpha \geq 0,
\end{equation}
since for the study of IR physics, the full features of the potential $V(\phi)$ in Eq.(\ref{V}) is not important for determining the low-energy behavior that will arise from the near-horizon region. However for our analysis of Seiberg--Witten stability, such an effective action will \emph{not} be suitable since branes are sensitive to the global geometry of the spacetime, after all we would like to know the sign of the Seiberg--Witten action at \emph{all} values of coordinate radius $r$, not just for the near-horizon regions.  

Finally we remark that charged dilaton AdS black holes that we are considering, since it is in $(4+1)$-dimensional spacetime, are only ``approximately'' asymptotically AdS, in the sense that the background metric (This is the background that one subtracts off in the Brown and York formalism to find the mass of the black hole) is not exactly $\text{AdS}_5$, though for large $r$, its asymptotic behavior is the same as that of $\text{AdS}_5$ (the background metric is exactly $\text{AdS}_n$ in higher dimension $n \geq 6$) \cite{HSD}. This however does not affect much of our discussions since Seiberg--Witten instability applies to any  spacetime (of dimension $\geq 4$) with an asymptotically hyperbolic Euclidean version (a Riemannian manifold is said to be asymptotically hyperbolic if it has a well-defined conformal boundary), even for string theory on $W^{d+1} \times Y^{9-d}$, where $W^{d+1}$ is an $(d+1)$-dimensional non-compact asymptotically hyperbolic manifold (generalizing $\text{AdS}_{n+1}$) and $Y^{9-d}$ is a compact space (generalizing $S^5$) \cite{McInnes5}.  The requirement that $W^{d+1}$ has a well-defined conformal boundary is somewhat crucial since it guarantees that the volume form $\omega$ on $W^{d+1}$ is exact, i.e. $\omega = d\mathcal{H}$ for some $d$-form $\mathcal{H}$, which physically is a field of the appropriate supergravity theory in $W^{d+1}$. In the Seiberg--Witten action [Eq.(\ref{SWaction})], If the probe brane $\Sigma$ is the boundary of a domain $\Omega$, then the volume enclosed by the brane is simply 
\begin{equation}
\text{Vol.} = \int_{\Omega} \omega = \int_{\Sigma} \mathcal{H},
\end{equation}
where the second equality follows from Stoke's theorem. In the case where the Seiberg--Witten action becomes negative, brane-anti-brane pairs are nucleated in close analogy to Schwinger pair-production, at the expense of the background $\mathcal{H}$ field \cite{Barbon}. 

We are now ready to compute the Seiberg--Witten action. We first Wick-rotate ($\tau = it$) our black hole metric to obtain the Euclidean version. The Seiberg--Witten action, as can be seen from Eq.(\ref{SWaction}), is in danger of becoming negative if the second term proportional to volume is large, and the most dangerous case is therefore when the charge $\mu$ attains its maximal value. This is the BPS case with $\mu=4\Theta$. Consequently, the Seiberg--Witten action is then 
\begin{equation}
S_{\text{SW}}= \Theta \int d\tau \sqrt{g_{\tau\tau}} \int d\Omega \left[f(r)\right]^3 - 4\Theta \int d\tau \int dr \int d\Omega \left[f(r)\right]^3\sqrt{g_{\tau \tau}}\sqrt{g_{rr}}. 
\end{equation} 
The ``time'' coordinate is now, as usual, identified periodically with some period $2\pi P$. For constant $r$, instead of $\Bbb{R} \times T^3$, we now have topology of a 4-torus $T^4$, which of course, is not necessarily (hyper)cubic (i.e. $K$ need not be the same as $P$). Pulling out all the common positive factors, we get
\begin{flalign}
S_{\text{SW}} \propto & ~~r^3\left[1-\left(\frac{b}{r}\right)^2\right]^{\frac{3\alpha^2}{2(2+\alpha^2)}}\left(-\left(\frac{c}{r}\right)^2\left[1-\left(\frac{b}{r}\right)^2\right]^{\frac{2-\alpha^2}{2+\alpha^2}} + r^2\left[1-\left(\frac{b}{r}\right)^2\right]^\frac{\alpha^2}{2+\alpha^2}\right)^\frac{1}{2} \notag \\
& - 4 \int_{r_+}^r dr' r'^3 \left[1-\left(\frac{b}{r'}\right)^2\right]^\frac{\alpha^2}{2+\alpha^2}.
\end{flalign}  
Unlike the case for Reissner--Nordstr\"om black hole, the computation is more involved because the volume integral is non-trivial, which is due to the fact that in this case $g_{\tau\tau}g_{rr} \neq 1$ \cite{Jacobson}. For the case $\alpha=1$, we recover the result obtained in \cite{yenchin}, in which it is shown that the black hole is in fact stable, i.e. the Seiberg--Witten action is always positive, regardless of the amount of electrical charges dropped into the black hole. For general $\alpha$, the calculation is further complicated by the fact that we cannot solve for the event horizon $r_+$ explicitly in terms of the metric parameters, and so the volume integral cannot be carried out straight-forwardly. 

Instead, we will normalize the event horizon such that 
\begin{equation}
r_+ = 1 ~~~\text{(Normalization Condition)}.
\end{equation}
Originally, the event horizon is determined by the metric parameters $M$, $K$, $Q$ and $\alpha$. Once we fixed the horizon at unity, there should be new constraints on these parameters so that at least some of the initially independent parameters are no longer independent. Indeed, we see that the event horizon should satisfy $g_{tt} = 0$, that is, from Eq.(\ref{gtt}),
\begin{equation}
r_+^2\left[1-\left(\frac{b}{r_+}\right)^2\right]^\frac{\alpha^2}{2+\alpha^2} = \left(\frac{c}{r_+}\right)^2\left[1-\left(\frac{b}{r_+}\right)^2\right]^{\frac{2-\alpha^2}{2+\alpha^2}}.
\end{equation}
Fixing $r_+=1$ means that 
\begin{equation}
(1-b^2)^{{}^{\frac{\alpha^2}{2+\alpha^2}}} = c^2\left(1-b^2\right)^{{}^\frac{2-\alpha^2}{2+\alpha^2}}.
\end{equation}
If $b^2 \neq 1$ (Recall from previous Section that $r=b$ is a curvature singularity, $r_+ \to b$ is the extremal limit), then we can solve for $c$ and obtain
\begin{equation}\label{bc-relation}
c=(1-b^2)^{{}^{\frac{\alpha^2-1}{\alpha^2+2}}} \Leftrightarrow b^2 = 1 - c^{{}^{\frac{2+\alpha^2}{\alpha^2-1}}}, ~~\alpha \neq 1.
\end{equation}  

The condition in Eq.(\ref{bc-relation}) is equivalent to, in terms of the metric parameters $Q$, $M$, $K$ and $\alpha$, 
\begin{equation}\label{pre-MQ}
1-\frac{Q^2(2+\alpha^2)}{64\pi^3K^6M} = \left(\frac{2M}{3\pi K^3}\right)^\frac{2+\alpha^2}{2\alpha^2-2}.
\end{equation}  
That is, we have the following result:

\textbf{MQ-Relation:} \emph{In our normalization choice of $r_+ =1$, the mass $M$ of the locally asymptotically anti-de Sitter charged dilaton black hole with flat event horizon is related to its electrical charge $Q$ by}
\begin{equation}\label{MQ-relation}
M\left[1-\left(\frac{2M}{3\pi^2 K^3}\right)^\frac{2+\alpha^2}{2\alpha^2-2}\right] = \frac{Q^2(2+\alpha^2)}{64\pi^3 K^6} \geq 0. 
\end{equation}

For any fixed $K$, this relation allows us to use $M$ as the charge parameter in our normalization. Using elementary calculus, we can show that the function 
\begin{equation}
f(M) = M\left[1-\left(\frac{2M}{3\pi^2 K^3}\right)^\frac{2+\alpha^2}{2\alpha^2-2}\right]
\end{equation}
is monotonically decreasing for all $0 < \alpha < 1$. Therefore we have the following result:

\textbf{Corollary 1}: \emph{For our choice of normalization $r_+ =1$, in the range of dilaton coupling $0 < \alpha <1$, increasing (resp. decreasing) the electrical charges $Q$ corresponds to increasing (resp. decreasing) the mass $M$, where $M\geq 3\pi^2K^3/2$.}

From Eq.(\ref{pre-MQ}) we note that $M=3\pi^2K^3/2$ is just the mass of the uncharged flat black holes under our renormalization condition.

For $\alpha >1$, the function $f(M)$ starts out at 0 when $M=0$, then tends to $-\infty$ as $M \to \infty$. It has a turning point at 
\begin{equation}
M_0 = M_0(\alpha) = \frac{3\pi^2 K^3}{2}\left(\frac{2\alpha^2-2}{3\alpha^2}\right)^\frac{2\alpha^2-2}{\alpha^2 + 2}.
\end{equation}
Note that as $\alpha \to \infty$, the turning point tends to a finite value $2\pi^2K^3/3$. Therefore we obtain

\textbf{Corollary 2}: \emph{For our choice of normalization $r_+ =1$, in the range of dilaton coupling $\alpha > 1$, increasing charges corresponds to decreasing mass, where $M$ satisfies}
\begin{equation}
\frac{3\pi^2 K^3}{2}\geq M> \frac{3\pi^2K^3}{2}\left(\frac{2\alpha^2-2}{3\alpha^2}\right)^\frac{2\alpha^2-2}{\alpha^2+2}.
\end{equation}

This echoes the finding that charged dilaton black holes in AdS behaves rather differently for $0<\alpha<1$ and $\alpha>1$ \cite{HSD} just like their simpler GHS cousin. 

Hendi, Sheykhi, and Dehghani also showed that the extremal radius $r_E$ of flat dilaton black holes in AdS satisfies \cite{HSD} 
\begin{equation}
r^2_E = \frac{3}{2+\alpha^2_E}b_E^2. 
\end{equation}
Every quantity with subscript $E$ means that said quantity is being evaluated at the value when the black hole is extremal.
With our normalization condition, this is equivalent to
\begin{equation}
b^2_E = \frac{2+\alpha^2_E}{3} = \frac{Q^2_E (2+\alpha^2_E)}{64\pi^3K^3M_E}.
\end{equation}
This implies
\begin{equation}\label{QE}
Q^2_E = \frac{64\pi^3K^3 M_E}{3}. 
\end{equation}

In previous work with $\alpha=1$, it was shown that without normalization condition, the extremal charge is given by \cite{yenchin}
\begin{equation}
Q_E (\alpha=1) = \frac{8\times 2^\frac{1}{4}}{3^\frac{3}{4}}\pi (KM)^{\frac{3}{4}}.
\end{equation}
For consistency, it can be easily checked that if we use the normalization condition, this result indeed agrees with Eq.(\ref{QE}).

In our choice of normalization, the Seiberg--Witten action becomes
\begin{flalign}
S_{\text{SW}} \propto & ~~r^3\left[1-\frac{1-c^{{}^{\frac{2+\alpha^2}{\alpha^2-1}}}}{r^2}\right]^{\frac{3\alpha^2}{2(2+\alpha^2)}}\left(-\left(\frac{c}{r}\right)^2\left[1-\frac{1-c^{{}^\frac{2+\alpha^2}{\alpha^2-1}}}{r^2}\right]^{\frac{2-\alpha^2}{2+\alpha^2}} + r^2\left[1-\frac{1-c^{{}^\frac{2+\alpha^2}{\alpha^2-1}}}{r^2}\right]^\frac{\alpha^2}{2+\alpha^2}\right)^\frac{1}{2} \notag \\
& - 4 \int_{1}^r dr' r'^3 \left[1-\frac{1-c^{{}^\frac{2+\alpha^2}{\alpha^2-1}}}{r'^2}\right]^\frac{\alpha^2}{2+\alpha^2},
\end{flalign}  
where $c^2=2M/(3\pi^2K^3)$.

\begin{center}
\begin{figure}
\includegraphics[width=5.5 in]{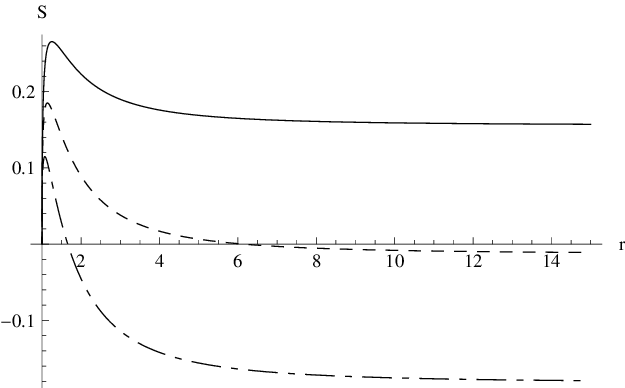}
\caption{For $\alpha=0.0001$, the Seiberg-Witten action behaves very much the same as the Reissner-Nordstr\"om case ($\alpha=0$): Increasing electrical charge lowers the action. Nevertheless, despite appearance the action actually does \emph{not} remain negative like its Reissner-Nordstr\"om counterpart; at sufficiently large $r$ the action turns around and eventually becomes positive. Here, as well as in all subsequent figures, solid line has the lowest amount of charges, while mixed-dash line has the highest amount of charges. Also, we have fixed $K=1$ in all figures.}
\label{fig:1}
\end{figure}
\end{center}

\begin{center}
\begin{figure}
\includegraphics[width=5.5 in]{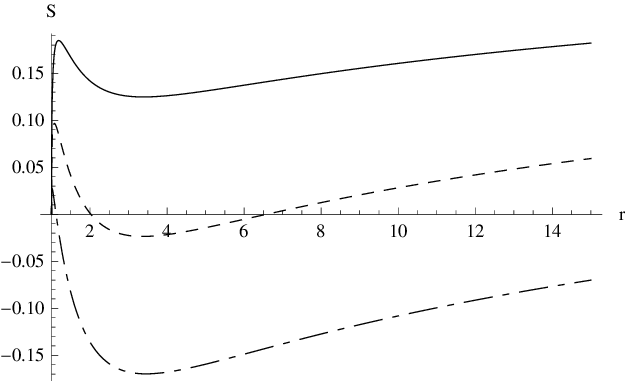}
\caption{For $\alpha=0.5$, the turn-around behavior is now obvious as it occurs at smaller value of $r$. In fact the radial coordinate of turning point decreases as $\alpha$ increases from 0 towards 1.}
\label{fig:2}
\end{figure}
\end{center}

\begin{center}
\begin{figure}
\includegraphics[width=5.5 in]{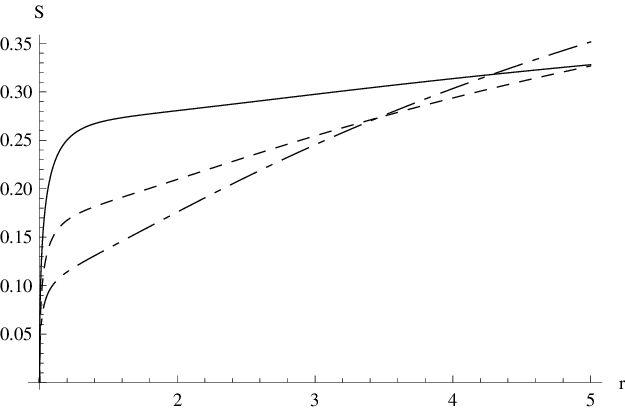}
\caption{At $\alpha=0.9$, the action is now very similar to the case $\alpha=1$ investigated in previous work \cite{yenchin}. For $\alpha \neq 0$, for typical fixed charge $Q_1$, increasing the charge to $Q_2 > Q_1$ makes the action starts out with $S(Q_2) < S(Q_1)$ initially, but subsequently takes over at some finite value $r=R$ so that $S(Q_2) > S(Q_1)$ for all $r \geq R$. The value of $R$ in which this takeover occurs decreases with increasing charges.}
\label{fig:3}
\end{figure}
\end{center}

We find that for $0<\alpha<1$, the action becomes lower as we increase the amount of electrical charges (equivalently, with our normalization, by increasing the mass). Just like its flat Reissner--Nordstr\"om counterpart, the action eventually becomes negative. However, the action does \emph{not} stay negative like the flat Reissner--Nordstr\"om case, as can been seen from Fig.(\ref{fig:1}), but instead turns around and eventually becomes positive again at large enough $r$. This can be seen numerically for sufficiently large value of $\alpha$ as in Fig.(\ref{fig:2}). We will provide an analytic proof for \emph{all} values of $\alpha$ in the appendix. The position of the turning point is numerically found to be dependent on $\alpha$: the closer $\alpha$ is to unity, the turning point occurs at smaller radial distance from the horizon. This is consistent with the previous result \cite{yenchin} that at $\alpha=1$ there is no turning point at all, and in fact the action never becomes negative. That is, the ``valley'' around the turning point becomes shallower as $\alpha \to 1$. As expected, the plot becomes similar to that of $\alpha=1$ case in this limit, as we can see by comparing Fig.(\ref{fig:3}) to Fig.1 of \cite{yenchin}.

\begin{center}
\begin{figure}
\includegraphics[width=5.5 in]{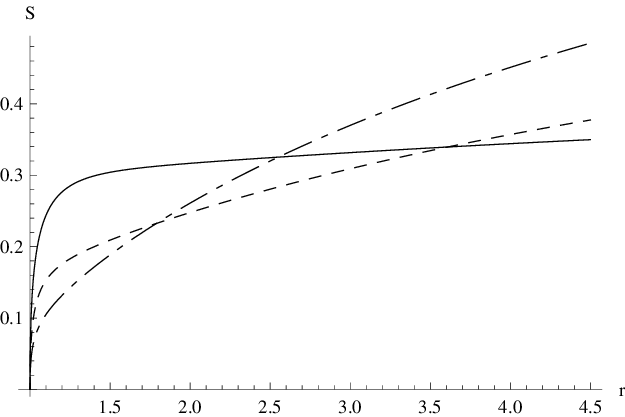}
\caption{As expected from continuity, similar behavior as in Fig.3 occurs for $\alpha=1.1$.}
\label{fig:4}
\end{figure}
\end{center}

\begin{center}
\begin{figure}
\includegraphics[width=5.5 in]{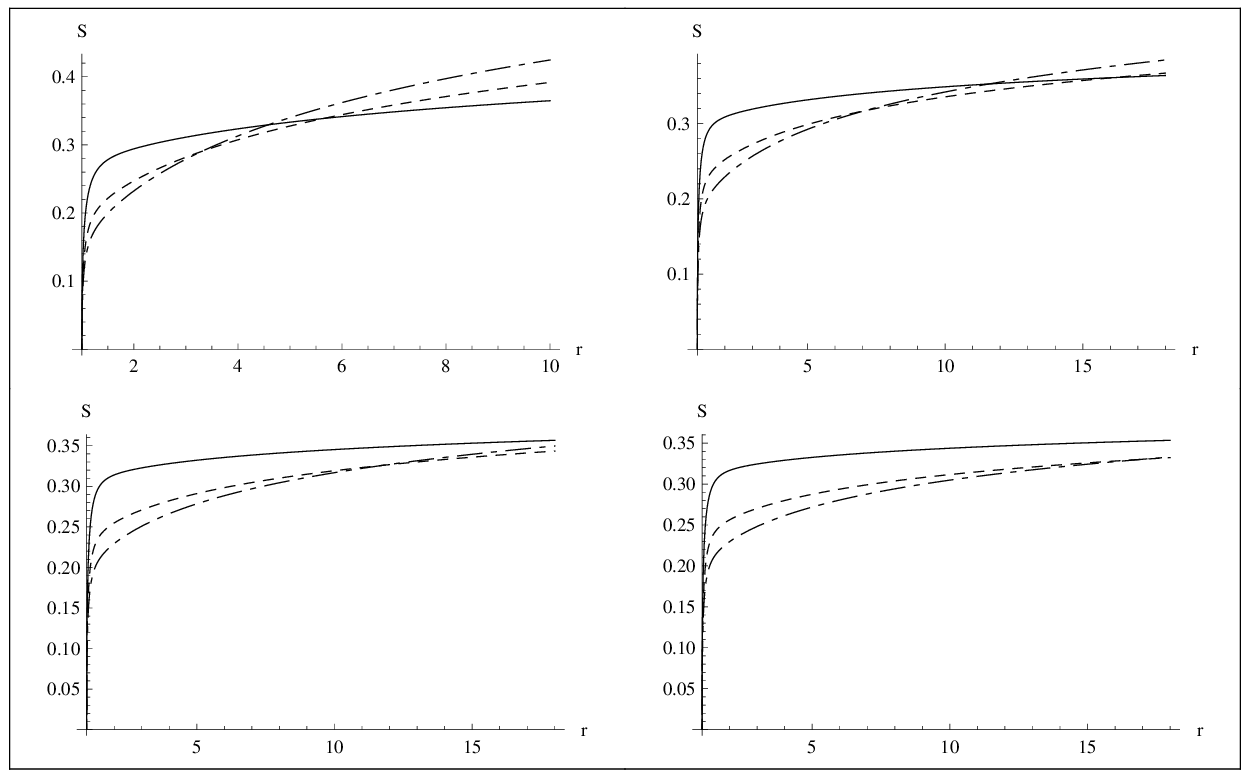}
\caption{For fixed electrical charge, increasing the value of $\alpha$ pushes the cross-over position in Fig.3 and Fig.4 further away from the horizon. Top-left, top-right, bottom-left, and bottom-right figures correspond to the choice $\alpha=2,~2.5,~2.8$ and $3$ respectively.}
\label{fig:5}
\end{figure}
\end{center}

For $\alpha > 1$, the behavior is similar to that of $\alpha=1$ case (Fig.(\ref{fig:4})) and the action remains positive for all values of admissible charges. For fixed electrical charge, increasing the value of $\alpha$ has the effect of pushing the cross-over position in Fig.(\ref{fig:3}) and Fig.(\ref{fig:4}) away from the horizon, as can be seen from Fig.(\ref{fig:5}).

\section{Discussion}
From the above analysis, it is straightforward to conclude that for $\alpha>1$, locally asymptotically AdS charged dilaton black holes with flat event horizon are stable in the Seiberg--Witten sense. The case $0 < \alpha <1$ requires more careful examination since even though for sufficiently high charges the action becomes negative, it does not stay negative but instead eventually climbs back up to positive values. This type of behavior was first pointed out and discussed by Maldacena and Maoz in the context of cosmology \cite{MaldacenaMaoz}.

One way to interpret this phenomenon is as follows: brane-anti-brane pairs are created at exponential rate from the reservoir of energy where the action is negative. For the action that is always positive, branes produced by pair-production (the production rate of which is suppressed by $\text{exp}(-S)$ anyway) can minimize the action by moving towards the horizon and collapsing to minimal (zero) size.  If the action becomes negative at large $r$ and \emph{remains} negative, the branes can then keep lowering the action by moving towards the boundary, growing indefinitely in the attempt to lower the action, signaling complete instability of the system.  For action which is only negative between some finite range, branes can minimize the action by moving to this region (Note that most of the brane-anti-brane pairs are actually created in this region in the first place due to the exponential enhancement in pair-production rate) instead of collapsing to zero size under their own tension. However the action of any brane can only be reduced by a \emph{finite} amount in this case. This leads Maldacena and Maoz to suspect that there should be ``nearby'' solutions that are stable. In a more dynamical view, if the action is only negative for some finite range of coordinate radius $r$, then the branes must be produced in such a way that some of the metric parameters of the black hole spacetime eventually be brought down below the threshold value that triggered the instability. However, when everything has settled down to a stable configuration, it is \textit{no longer the original spacetime}. It has become a ``nearby'' solution in the sense of Maldacena and Maoz. To put it in a slightly different, more physical language, a physical instability can never truly ``run away'' in general. The brane-anti-brane pairs will soon occupy the surrounding black hole environment due to the exponential rate of pair-production. This will likely alter the original boundary condition of the original action. As a result, the exponential pair-production will stop. A good analogy is the ionization, or break-down, of neutral hydrogen gas into plasma (we could say, ``pair-production'' of electrons and protons) under external $E$-field, say, between parallel plates. When the $E$-field reaches the atom's ionization energy within one \AA, there is an exponential avalanche. However this run away situation is quickly suppressed since the surrounding plasma would induce negative $E$-field to counter the original $E$-field. That is, instability is often \emph{self-limiting}. Having said that, we remark that Seiberg-Witten instability is \emph{not} always self-limiting. This is the consequence of the fact that \emph{there exist compact manifolds on which it is impossible to define a Riemannian metric of positive or zero scalar curvature} \cite{McInnes5}. For such cases, the AdS bulk is unstable due to emission of brane-anti-brane pairs and will \emph{remain unstable no matter how the metric is distorted due to backreaction}. This is the case for black holes with \emph{negatively} curved horizon in Einstein-Maxwell theory: once brane-anti-brane pairs are produced, nothing can stop the instability; no matter how the branes deform the spacetime, the scalar curvature at infinity can never become everywhere positive or zero. However, the situation is much more subtle with black hole solutions that have only finite ``negative reservoir''.

Explicit examples of black hole spacetimes with Seiberg--Witten action that only becomes negative for finite range of $r$ were previously found in the context of Ho\v{r}ava-Lifshitz gravity \cite{yenchin2}, in which it was pointed out that we are lacking a way to investigate this kind of instability quantitatively. \emph{Qualitatively} we expect that the more ``negative reservoir'' the action has become, the more unstable it is, in the sense that ``nearby'' solutions may not even exist. However since Ho\v{r}ava-Lifshitz gravity is \emph{not} a string theory, it is doubtful if naive application of Seiberg--Witten stability makes sense. Even if it does, one may reasonably suspect that many of the properties of the Seiberg--Witten action for Ho\v{r}ava-Lifshitz black holes are due mainly to the features of Ho\v{r}ava-Lifshitz gravity itself. In this work, we showed that Seiberg--Witten action for black hole spacetime can indeed become negative for finite range of coordinate radius $r$ only, with \emph{explicit} example of charged flat black hole in the context of Einstein-Maxwell-Dilaton theory, which \emph{can} be embedded in low energy string theory. In this case, since we know that flat Reissner--Nordstr\"om black holes ($\alpha=0$) become unstable at sufficiently large electrical charges, we should expect that for $\alpha$ very close to zero, the solutions, being a very small perturbation away from flat Reissner--Nordstr\"om solution, should also be unstable. It is then not surprising that we find exactly in this case a very large range of coordinate radius $r$ in which the Seiberg Witten action is negative.

Therefore we can conclude that for dilaton coupling $\alpha \approx 0$ the black hole is indeed unstable against uncontrolled brane nucleation, 
while for $\alpha>1$ the black hole spacetime is completely stable against brane nucleation. For $0 < \alpha < 1$ bounded away from zero, there might or might not exist ``near-by'' solutions in the sense of Maldacena and Maoz. 

We now remark on the field theory side of the story. In most applications of AdS/CMT, one typically considers a Reissner--Nordstr\"om black hole that eventually becomes unstable at sufficiently low temperature and is replaced by a hairy black hole (not necessary of dilaton type), see e.g. \cite{CDP, Martinez, Kolyvaris}. Dually this corresponds to condensate of some kind. For its potential applications to condensed matter systems, while it is good that we have found that flat charged dilaton black holes are stable in the Seiberg--Witten sense, at least for some values of the dilaton coupling $\alpha$ for \emph{any} amount of admissible charge, we might reasonably worry about whether the zero temperature limit is well-behaved. As pointed out by Hartnoll and Tavanfar \cite{HT}, if a black hole is present in the extremal limit, which corresponds to zero temperature on the dual field theory, then we do not have explicit access to the zero temperature degree of freedom since in the gravity side the horizon shielded some of the information. That is, as Hartnoll and Tavanfar put it, a black hole is problematic precisely because of their ``blackness''. Therefore a black hole description is entirely inappropriate for dual field theories at low temperatures. We should therefore address the (apparent) stability of some of these dilaton black holes.

We first remind the readers that even if a black hole spacetime is stable in Seiberg--Witten sense, this is far from a guarantee that they are indeed stable. As pointed out by Hartnoll and Tavanfar, there are a myriad of ways that a Reissner--Nordstr\"om black hole can become unstable. Even in the case of charged dilaton black holes, in principle one has to check each and every type of instability before we can conclude that a specific black hole solution is indeed stable. Indeed, it is interesting to contrast our result with the \emph{thermodynamic} analysis performed by Hendi, Sheykhi, and Dehghani \cite{HSD}. The Hawking temperature approaches a constant value as $r_+ \to b$ in the extremal limit. They also showed that for $\alpha > 1$, large ($r_+ \gg b$) charged dilaton black holes in AdS are thermodynamically stable for sufficiently large value of $\alpha$, whereas small ones ($r_+ \ll b$) are thermodynamically unstable for some range of $\alpha$. 

Even \emph{if} there really does exist some values of $\alpha$ that admit charged dilaton black hole solutions that are completely stable against all types of instability for all values of charges, we see that there is a natural way to evade the worry raised by Hartnoll and Tavanfar: the flat charged dilaton black holes do not achieve zero temperature at extremal limit (so that they naturally do not describe a system at very low temperatures), and furthermore they \emph{do not have event horizon} in the extremal limit: they ``cease to be black'' and thus do not conceal any degree of freedom. Instead, the extremal limit is that of null singularity. From various studies on time-dependent AdS/CFT duality in the context of cosmology, we know that not all naked singularities lead to problems on the dual field theories \cite{Das1,Ho,Das2}. In other words, naked singularities are not necessarily disallowed in AdS/CFT. Gubser has formulated some criteria for naked singularities to be acceptable in the AdS bulk \cite{Gubser3}. It remains of course to be explicitly checked if these singularities resulting from dilation black holes are allowable. Even in the case of charged dilaton black holes with \emph{positively curved} event horizon, for which an extremal black hole does have zero entropy \cite{CDP, SDH}, they are not really black hole since their horizon degenerates. See also \cite{HoRo} for related discussion. Thus in any case we do not expect charged dilaton black holes to describe field theories at zero temperature limit, so the worry of Hartnoll and Tavanfar does not arise. This is of course, subjected to the caveat that [in the magnetic charged case] due to Eq.(\ref{stringframe}), string coupling near extremal horizon becomes very large, thus signaling the failure of semi-classical treatment of gravity. Higher order corrections to gravity becomes necessary if we want to be confident about the properties of extremal [magnetically] charged dilaton black holes. [Strictly speaking these statements are about the asymptotically flat case, and should be checked explicitly for asymptotically AdS black holes.] It could be possible that in the full theory singularities do not arise, and that in the extremal limit the charged dilaton black holes make a transition into another geometry without event horizon, instead of a null singularity. See also \cite{Townsend} and \cite{Cox} for related discussion on singularity removal of charged dilaton black holes. 

We now comment on further work to be done in this area. Firstly, since magnetic field is an important parameter in many condensed matter system, it would be interesting to study also the Seiberg--Witten stability for charged dyonic dilaton black holes which are asymptotically AdS. These black holes are solutions to Einstein-Maxwell theory coupled to a dilaton-axion field. While holography of such black holes have already been studied, (see. e.g. \cite{GLKPTW}), complication may arise since dyonic dilaton black hole solutions cannot be embedded into supergravity theory except for $\alpha=0,\sqrt{3}$ \cite{Nozawa}. Also of importance is, as we have just remarked, higher order corrections to gravity is important to understand the extremal magnetically charged dilaton black holes [at least in the asymptotically flat case with spherical topology]. Therefore it might be of relevance to consider for example asymptotically AdS charged dilaton black holes with the Gauss-Bonnet term \cite{GOT, OT} and study their stability. In addition, in \cite{fbh}, it was pointed out that a squashed toral black hole is extremely \emph{fragile}, a term coined in \cite{fbh} to describe the fact that even the tiniest non-trivial deformations of the squashed flat black hole leads to instability in Seiberg-Witten sense, it would be interesting to consider the effect of dilaton on the fragility of these black holes. Another natural generalization to consider would be to extend the present work to arbitrary spacetime dimension more than or equal to 4. 

Lastly, it is important to further understand Seiberg--Witten instability in the case where the action is only negative for finite range of coordinate radius $r$. As discussed at the beginning of this Section, presumably the emission of branes will back-react on the spacetime metric. In most contexts of cosmology \cite{MaldacenaMaoz}, even finite amount of ``negative reservoir'' for the Seiberg--Witten action is very severe because in cosmology spacetime cannot ``change into something else'' whereas black holes can, e.g. by losing mass, charge, angular momentum, etc. and eventually settle down to a ``nearby solution''. Therefore, it is important to investigate how exactly this is achieved, and to determine the conditions for ``nearby solutions'' to exist. However we do not know how to do so quantitatively. The fact that this type of behavior can arise for relatively simple black hole spacetime in Einstein-Maxwell-Dilaton theory makes these questions even more pressing to resolve. We will return to this question in future work. Finally we should discuss robustness of our present work.

\subsection{Robustness of Results}

In our analysis thus far, we only considered Gao-Zhang solution of charged dilaton AdS black hole. This corresponds to the choice of a very special potential of the form in Eq.(\ref{V}). However other types of potentials have been considered in various works, e.g. \cite{Cadoni}. As mentioned in Section 3 however, it is important that one uses the \emph{full} potential that works \emph{globally} in the bulk, and not an approximate one that only works close to the horizon of the black hole. The latter suffices for most applications of AdS/CMT, but \emph{not} for Seiberg-Witten stability analysis, since typically the Seiberg-Witten brane action becomes negative not at the vicinity of, but some distance away, from the horizon. In other words, without specifying the \emph{full profile} of the non-Gao-Zhang type potentials, they are not readily subjected to the stability analysis that we have carried out in this work. Nevertheless, we can still deduce some results about the robustness of our analysis, assuming that there indeed exists full profile for the dilaton field that are not of Gao-Zhang type, which nevertheless gives rise to well-defined conformal boundary.

Recall that the potential in Eq.(\ref{V}) is specially tailored so that the black hole is locally asymptotically anti-de Sitter. One could imagine starting with such a potential, but gradually deform it continuously from its exact form in Eq.(\ref{V}). This would result in different black hole solutions in the bulk and correspondingly the distortion of the geometry of the conformal boundary (but not its topology since we are only making continuous deformation), i.e. the torus at the boundary is prescribed with different metric depending on the choice of the potentials. This is guaranteed by Anderson-Chrusciel-Delay theorem \cite{ACD}, which basically says that if one continuously deforms the torus at infinity for the Euclidean flat AdS black hole, so that it remains ``static'' but ceases to be flat, then one can still find a ``static'' Euclidean AdS black hole which induces said deformed metric at infinity (This can however fail if the deformation is too extreme). However, provided that the scalar curvature of the distorted torus at infinity is not exactly zero, its metric will always gives rise to negative scalar curvature after conformal transformation. This is a consequence of deep result in global differential geometry, from the work of Schoen and Yau \cite{SY}, as well as Gromov and Lawson \cite{GL}. The work of Seiberg and Witten however connects negative scalar curvature at conformal boundary with instability induced by tachyonic mode on the field theory \cite{SeibergWitten}. 

In other words, distorting the Gao-Zhang solution by deforming the potential generically leads to instability. Thus for small dilaton coupling $\alpha \approx 0$, sufficiently charged dilaton black holes that are unstable from our analysis will continue to be unstable under such deformation. The case of vanishing scalar curvature at conformal infinity of course requires more careful, separate investigations, perhaps in the line of \cite{McInnes0120}. Similar result holds for the case $0 < \alpha < 1$. In other words, while one expects the quantitative details to change when we continuously deform the potential, and hence the black hole solution, the qualitative picture should be expected to hold as long as the new solutions are not too drastically different from the original Gao-Zhang solution. This argues that \emph{as far as instability is concerned}, our result should be \emph{qualitatively} robust. 

Note, however, that by the same token, one should expect that even for strong dilaton coupling $\alpha > 1$, arbitrary deformation should lead to generic \emph{instability}, so the Seiberg-Witten \emph{stability} of the Gao-Zhang black hole for $\alpha > 1$ might indeed be special and might not hold when the potential is changed. The scalar-flat case should again be checked explicitly, which we propose to discuss elsewhere in future work.

\newpage
\section{Appendix: Details of Proofs}

\textbf{Theorem 1:} \emph{The Seiberg-Witten action of a locally asymptotically anti-de Sitter charged dilaton Gao-Zhang black hole with flat event horizon in 5 dimension is logarithmically divergent at large $r$ limit for finite values of $\alpha$.}

\begin{proof} We recall that to obtain the Euclidean metric, we Wick-rotate so that $t$ now parametrizes a 4-th circle, in addition to the 3 circles that parametrize the flat torus. The Euclidean metric is thus a metric on a manifold which is radially foliated by copies of the 4-torus.
One can think of $t/L$  as an angular coordinate on this 4-th circle with periodicity $2\pi P$ chosen so that the metric is not singular at the horizon $r_+$. This gives rise to conformal boundary which is a conformal torus. The field theory is then defined on this space which is the Euclidean version of ordinary flat spacetime with periodic boundary conditions. 

The Seiberg-Witten action is 
\begin{equation}
S_{\text{SW}}= \Theta \int d\tau \sqrt{g_{\tau\tau}} \int d\Omega \left[f(r)\right]^3 - 4\Theta \int d\tau \int dr \int d\Omega \left[f(r)\right]^3\sqrt{g_{\tau \tau}}\sqrt{g_{rr}}. 
\end{equation} 
Explicitly, with our normalization condition $r_+=1$, we have
\begin{flalign}
\frac{S_{\text{SW}}}{2\pi PL\Theta \Gamma_0} &= r^3\left[1-\frac{1-c^{{}^{\frac{2+\alpha^2}{\alpha^2-1}}}}{r^2}\right]^{\frac{3\alpha^2}{2(2+\alpha^2)}} \left(-\left(\frac{c}{r}\right)^2\left[1-\frac{1-c^{{}^\frac{2+\alpha^2}{\alpha^2-1}}}{r^2}\right]^{\frac{2-\alpha^2}{2+\alpha^2}} + r^2\left[1-\frac{1-c^{{}^\frac{2+\alpha^2}{\alpha^2-1}}}{r^2}\right]^\frac{\alpha^2}{2+\alpha^2}\right)^\frac{1}{2} \notag \\ &- 4 \int_{1}^r dr' r'^3 \left[1-\frac{1-c^{{}^\frac{2+\alpha^2}{\alpha^2-1}}}{r'^2}\right]^\frac{\alpha^2}{2+\alpha^2},
\end{flalign}  
where $c^2=2M/(3\pi^2K^3)$, and $\Gamma=8\pi^3K^3$ is the area of a 3-torus.
We have restored the proportionality constant $2\pi PL\Theta \Gamma_0$ for completeness sake. We did not include this factor previously for simplicity (since we are only interested at where the action becomes negative, this \emph{positive} constant overall factor is of no importance).

Assuming first that $\alpha$ is finite. The first term is, with $A(\alpha):=1-c^{{}^\frac{2+\alpha^2}{\alpha^2-1}}$, 
\begin{flalign}
&r^3\left[1-\frac{A(\alpha)}{r^2}\right]^{\frac{3\alpha^2}{2(2+\alpha^2)}}\left[-\left(\frac{c}{r}\right)^2\left[1-\frac{A(\alpha)}{r^2}\right]^{\frac{2-\alpha^2}{2+\alpha^2}} +r^2\left[1-\frac{A(\alpha)}{r^2}\right]^{\frac{\alpha^2}{2+\alpha^2}}\right]^{\frac{1}{2}}\\
&=r^3\left[1-\frac{A(\alpha)}{r^2}\right]^{\frac{3\alpha^2}{2(2+\alpha^2)}}r\left[1-\frac{A(\alpha)}{r^2}\right]^{\frac{\alpha^2}{2(2+\alpha^2)}}\left[1-\left(\frac{c}{r}\right)^2\frac{1}{r^2}\left[1-\frac{A(\alpha)}{r^2}\right]^{\frac{2-2\alpha^2}{2+\alpha^2}}\right]^{\frac{1}{2}}\\
&=r^4\left[1-\frac{A(\alpha)}{r^2}\right]^{\frac{2\alpha^2}{2+\alpha^2}}\left[1-\frac{c^2}{2r^4}\left(1-\left(\frac{A(\alpha)}{r^2}\right)\left(\frac{2-2\alpha^2}{2+\alpha^2}\right)\right)+O(r^{-8})\right]\\
&=r^4\left[1-\frac{A(\alpha)}{r^2}\right]^{\frac{2\alpha^2}{2+\alpha^2}}\left[1-\frac{c^2}{2r^4} + O(r^{-6})\right]\\ 
&=\left(r^4-\frac{c^2}{2} + O(r^{-2})\right)\left[1-\frac{2\alpha^2}{2+\alpha^2}\frac{A(\alpha)}{r^2} + \left(\frac{1}{2}\right)\left(\frac{2\alpha^2}{2+\alpha^2}\right)
\left(\frac{\alpha^2-2}{2+\alpha^2}\right)\left(\frac{A^2(\alpha)}{r^4}\right) + O(r^{-6})\right] \\
&=r^4 -\frac{2\alpha^2}{2+\alpha^2}A(\alpha)r^2 + \left(\frac{\alpha^2}{2+\alpha^2}\right)\left(\frac{\alpha^2-2}{2+\alpha^2}\right)A^2(\alpha) - \frac{c^2}{2} + O(r^{-2}). \label{one}
\end{flalign}

The second term yields
\begin{flalign}
&4\int_{1}^{r} dr'^3 r'^3 \left[1- \frac{A(\alpha)}{r'^2}\right]^{\frac{\alpha^2}{2+\alpha^2}}\\
&=4\int_{1}^{r} dr' r'^3\left[1-\frac{A(\alpha)}{r'^2}\left(\frac{\alpha^2}{2+\alpha^2}\right)+\left(\frac{1}{2}\right)\left(\frac{\alpha^2}{2+\alpha^2}\right)\left(\frac{-2}{2+\alpha^2}\right)\frac{A^2(\alpha)}{r'^4} + O(r'^{-6})\right]\\
& = 4\int_1^r dr' \left[r'^3-\frac{A(\alpha)\alpha^2}{2+\alpha^2}r' -\left(\frac{\alpha^2}{(2+\alpha^2)^2}\right)\frac{A^2(\alpha)}{r'} + O(r'^{-3})\right] \\
& = 4\Bigg[\frac{r'^4}{4}-\frac{A(\alpha)\alpha^2}{2+\alpha^2}\left(\frac{r'^2}{2}\right)-\frac{\alpha^2}{(2+\alpha^2)^2}A^2(\alpha)\ln{r'}\Bigg]_1^r + O(1)\\
& = \left[r^4 - \frac{2A(\alpha)\alpha^2}{2+\alpha^2}r^2 - \frac{4\alpha^2}{(2+\alpha^2)^2}A^2(\alpha)\ln{r}\right]-\left[1-\frac{2A(\alpha)\alpha^2}{(2+\alpha^2)}\right] + O(1). \label{two}
\end{flalign}

Therefore, Eq.(\ref{one}) minus Eq.(\ref{two}) yields 
\begin{equation}\label{three}
\left(\frac{\alpha^2}{2+\alpha^2}\right)\left(\frac{\alpha^2-2}{2+\alpha^2}\right)A^2(\alpha) - \frac{2A(\alpha)\alpha^2}{2+\alpha^2}+1-\frac{c^2}{2} + \frac{4\alpha^2}{(2+\alpha^2)^2}A^2(\alpha)\ln{r} + O(1).
\end{equation}

The first four terms are also of order $O(1)$. Thus, in the limit of large $r$ and finite $\alpha$, 
\begin{equation}
S_{\text{SW}} \sim  2\pi PL\Theta \Gamma_0 \left[\frac{4\alpha^2}{(2+\alpha^2)^2}A^2(\alpha)\ln{r}\right] + O(1)
\end{equation}
which diverges as $r \to \infty$.
\end{proof}

Note that for $\alpha=0$, the last $O(1)$ term in Eq.(\ref{three}) also vanishes. This implies
\begin{equation}\label{four}
\lim_{r \to \infty} S_{\text{SW}} = 2\pi PL\Theta \Gamma_0 \left(1-\frac{c^2}{2}\right) = 2\pi PL\Theta \Gamma_0\left(1-\frac{\frac{2M}{3\pi^2K^3}}{2}\right)=\frac{1}{2}
\end{equation}
since for $\alpha=0$, we recall that from Eq.(\ref{pre-MQ}) the mass of flat uncharged black hole is, with our normalization condition, $M=3\pi^2K^3/2$.
For consistency check, we note that the flat uncharged black hole has metric \cite{McInnes3}
\begin{equation}
ds^2=-\left[\frac{r^2}{L^2}-\frac{2M}{3\pi^2K^3r^2}\right]dt^2 + \left[\frac{r^2}{L^2}-\frac{2M}{3\pi^2K^3r^2}\right]^{-1}dr^2+ r^2d\Omega_{k=0}^2,
\end{equation} 
and so its Seiberg-Witten action is
\begin{flalign}
S_{\text{SW}}(Q=0) &= 2\pi PL\Theta \Gamma_0 \left[r^3\left(\frac{r^2}{L^2}-\frac{16\pi M}{3\Gamma_0 r^2}\right)^\frac{1}{2}-\frac{r^4-r_+^4}{L}\right] \\
&= 2\pi PL\Theta \Gamma_0 \left[\frac{L\left(-\frac{16\pi M}{3\Gamma_0}\right)}{1+\left[1-\frac{16\pi M L^2}{3 \Gamma_0 r^4}\right]^{\frac{1}{2}}} + \frac{r_+^4}{L}\right]. \\
\end{flalign}
With the horizon easily computed to be
\begin{equation}
r_+ = \left(\frac{2ML^2}{3\pi^2K^3}\right)^\frac{1}{4},
\end{equation}
we have
\begin{equation}
\lim_{r \to \infty} S_{SW}(Q=0) = 2\pi PL\Theta \Gamma_0 \left[\frac{ML}{3\pi^2K^3}\right]. 
\end{equation}
Thus with our normalization condition, and $L=1$, we get indeed Eq.(\ref{four}). 

In the limit $\alpha \to \infty$ however, the Seiberg-Witten action remains \emph{finite} though still positive as $r$ tends to infinity:

\textbf{Theorem 2:} \emph{The Seiberg-Witten action of a locally asymptotically anti-de Sitter charged dilaton Gao-Zhang black hole with flat event horizon in 5 dimension in the limit $\alpha, r \to \infty$ is $S_{\text{SW}}=M/(3\pi K^3)$}. 

\begin{proof}
Denote $\displaystyle \lim_{\alpha \to \infty} A(\alpha) := A = 1- c$. The Seiberg-Witten action in the limit $\alpha \to \infty$ is
\begin{equation}
r^3\left[1-\frac{A}{r^2}\right]^{\frac{3}{2}}\left[-\frac{c^2}{r^2}\left(1-\frac{A}{r^2}\right)^{-1} + r^2\left(1-\frac{A}{r^2}\right)\right]^{\frac{1}{2}} - 4\int_1^r dr' r'^3\left(1-\frac{A}{r'^2}\right).
\end{equation}
The first term is 
\begin{flalign}
&r^3\left[1-\frac{A}{r^2}\right]^{\frac{3}{2}}\left[-\frac{c^2}{r^2}\left(1-\frac{A}{r^2}\right)^{-1} + r^2\left(1-\frac{A}{r^2}\right)\right]^{\frac{1}{2}}\\
&=r^3\left[1-\frac{A}{r^2}\right]^{\frac{3}{2}}\left\{r\left(1-\frac{A}{r^2}\right)^{\frac{1}{2}}\left[-\frac{c^2}{r^4}\left(1-\frac{A}{r^2}\right)^{-2}+1\right]^{\frac{1}{2}}\right\}\\
&=r^4\left[1-\frac{A}{r^2}\right]^2\left\{1-\frac{c^2}{2r^4}\left(1+\frac{2A}{r^2}+O(r^{-4})\right)\right\}\\
&=\left(1-\frac{2A}{r^2}+\frac{A^2}{r^4}\right)\left\{r^4-\frac{c^2}{2}\left[1+\frac{2A}{r^2}+O(r^{-4})\right]\right\}\\
&=r^4 - 2Ar^2 + A^2 - \frac{c^2}{2} + O(r^{-2}) \label{onealpha}.
\end{flalign}
The second term is
\begin{equation}
4\int_1^r dr' r'^3\left(1-\frac{A}{r'^2}\right)=r^4 - 2r^2A - 1 + 2A \label{twoalpha}.
\end{equation}
Thus Eq.(\ref{onealpha}) minus Eq.(\ref{twoalpha}) results in, as $r \to \infty$,
\begin{flalign}
A^2 -\frac{c^2}{2} + 1 - 2A &=(1-c)^2 -\frac{c^2}{2} + 1 - 2(1-c) \\
&=\frac{c^2}{2} = \frac{M}{3\pi^2K^3},
\end{flalign}
as desired.
\end{proof}

\acknowledgments
The authors would like to thank Brett McInnes from National University of Singapore for his useful comments. Yen Chin Ong would also like to thank Fech Scen Khoo, a much appreciated comrade, for checking part of the calculations. Pisin Chen is supported by Taiwan National Science Council under Project No. NSC 97-2112-M-002-026-MY3, by Taiwan's National Center for Theoretical Sciences (NCTS), and by US Department of Energy under Contract No. DE-AC03-76SF00515. Yen Chin Ong is supported by the Taiwan Scholarship from Taiwan's Ministry of Education.

\end{document}